\documentclass[twocolumn]{aastex63}

\usepackage{amsmath}
\usepackage{CJK}
\usepackage{float}
\usepackage[normalem]{ulem}
\usepackage[shortlabels]{enumitem}
\usepackage{multirow}
\usepackage{hyperref}
\definecolor{myred}{RGB}{255, 50, 50} 
\definecolor{myblue}{RGB}{0, 0, 180} 
\definecolor{mygrey}{RGB}{184, 134, 11} 
\definecolor{green}{RGB}{0, 180, 0}


\def\h  {\ifmmode{^{\rm h}}\else{$^{\rm h}$}\fi}
\def\m  {\ifmmode{^{\rm m}}\else{$^{\rm m}$}\fi}
\def\s  {\ifmmode{^{\rm s}}\else{$^{\rm s}$}\fi}
\def\deg  {\ifmmode{^{\circ}}\else{$^{\circ}$}\fi}
\def\decdeg {\ifmmode{{\rlap.}^{\circ}} \else ${\rlap.}^{\circ}$\fi}
\def\decs {\ifmmode{{\rlap.}^{\rm s}} \else ${\rlap.}^{\rm s}$\fi}
\def\decas {\ifmmode{{\rlap.}{''}}\else{${\rlap.}{''}$}\fi}
\def\Ho {\ifmmode{{\rm H}_0}\else{H$_0$}\fi}
\def\Ro {\ifmmode{{\rm R}_0}\else{R$_0$}\fi}
\def\To {\ifmmode{\Theta_0}\else{$\Theta_0$}\fi}

\def\Vsbar {\ifmmode {\overline{V_s}}\else {$\overline{V_s}$}\fi}
\def\Usbar {\ifmmode {\overline{U_s}}\else {$\overline{U_s}$}\fi}
\def\Wsbar {\ifmmode {\overline{W_s}}\else {$\overline{W_s}$}\fi}

\def\mux {\ifmmode {\mu_x}\else {$\mu_x$}\fi}
\def\muy {\ifmmode {\mu_y}\else {$\mu_y$}\fi}
\def\mura {\ifmmode {\mu_{\alpha}}\else {$\mu_{\alpha}$}\fi}
\def\mude {\ifmmode {\mu_{\delta}}\else {$\mu_{\delta}$}\fi}

\def\gax{\mathrel{\rlap{\lower4pt\hbox{\hskip1pt$\sim$}}
		\raise1pt\hbox{$>$}}}

\def\d {\ifmmode {{\rlap{.}}^\circ}\else {${\rlap{.}}^\circ$}\fi}
\def\s {\ifmmode {{\rlap{.}}^s}\else {${\rlap{.}}^s$}\fi}
\def\as {\ifmmode {{\rlap{.}}^{''}}\else {${\rlap{.}}^{''}$}\fi}

\def\lax{\mathrel{\rlap{\lower4pt\hbox{\hskip1pt$\sim$}}
		\raise1pt\hbox{$<$}}}    % less than or approx. symbol
\def\gax{\mathrel{\rlap{\lower4pt\hbox{\hskip1pt$\sim$}}
		\raise1pt\hbox{$>$}}}    % greater than or approx. symbol

\def\aap{A\&A}

\def\apjs{ApJS}
\def\apjl{ApJ}

\shorttitle{Diffuse Ionized Gas in the Anti-center of the Milky Way}
\shortauthors{Wen et al.}
\graphicspath{{./}{./Figure/}}

\begin{document}

\title{Diffuse Ionized Gas in the Anti-center of the Milky Way}

\author[0009-0008-1361-4825]{Shiming Wen} 
\affiliation{CAS Key Laboratory of Optical Astronomy, National Astronomical Observatories, Chinese Academy of Sciences, Beijing 100101, People's Republic of China; xtwfn@bao.ac.cn}

\author[0000-0002-1783-957X]{Wei Zhang}
\affiliation{CAS Key Laboratory of Optical Astronomy, National Astronomical Observatories, Chinese Academy of Sciences, Beijing 100101, People's Republic of China; xtwfn@bao.ac.cn}

\author{Lin Ma}
\affiliation{{The Key Laboratory of Cosmic Rays (Tibet University), Ministry of Education, Lhasa 850000, Tibet, China}}
\affiliation{CAS Key Laboratory of Optical Astronomy, National Astronomical Observatories, Chinese Academy of Sciences, Beijing 100101, People's Republic of China; xtwfn@bao.ac.cn}
\affiliation{School of Astronomy and Space Science, University of Chinese Academy of Sciences, Beijing 100049, China}

\author[0009-0000-6954-9825]{Yunning Zhao}
\affiliation{CAS Key Laboratory of Optical Astronomy, National Astronomical Observatories, Chinese Academy of Sciences, Beijing 100101, People's Republic of China; xtwfn@bao.ac.cn}
\affiliation{School of Astronomy and Space Science, University of Chinese Academy of Sciences, Beijing 100049, China}

\author{Lam, Man I.}
\affiliation{CAS Key Laboratory of Space Astronomy and Technology, National Astronomical Observatories, Chinese Academy of Sciences, Beijing 100101, People's Republic of China}

\author{Chaojian Wu}
\affiliation{CAS Key Laboratory of Optical Astronomy, National Astronomical Observatories, Chinese Academy of Sciences, Beijing 100101, People's Republic of China; xtwfn@bao.ac.cn}

\author{Juanjuan Ren}
\affiliation{CAS Key Laboratory of Space Astronomy and Technology, National Astronomical Observatories, Chinese Academy of Sciences, Beijing 100101, People's Republic of China}

\author{Jianjun Chen}
\affiliation{CAS Key Laboratory of Space Astronomy and Technology, National Astronomical Observatories, Chinese Academy of Sciences, Beijing 100101, People's Republic of China}

\author{Yuzhong Wu}
\affiliation{CAS Key Laboratory of Optical Astronomy, National Astronomical Observatories, Chinese Academy of Sciences, Beijing 100101, People's Republic of China; xtwfn@bao.ac.cn}

\author[0000-0003-1828-5318]{Guozhen Hu}
\affiliation{College of Physics, Hebei Normal University, 20 South Erhuan Road, Shijiazhuang 050024, People's Republic of China}

\author{Yonghui Hou}  
\affiliation{Nanjing Institute of Astronomical Optics, \& Technology, National Astronomical Observatories, Chinese Academy of Sciences, Nanjing 210042, People's Republic of China}
\affiliation{School of Astronomy and Space Science, University of Chinese Academy of Sciences, Beijing 100049, China}

\author{Yongheng Zhao}
\affiliation{CAS Key Laboratory of Optical Astronomy, National Astronomical Observatories, Chinese Academy of Sciences, Beijing 100101, People's Republic of China; xtwfn@bao.ac.cn}
\affiliation{School of Astronomy and Space Science, University of Chinese Academy of Sciences, Beijing 100049, China}

\author{Hong Wu}
\affiliation{CAS Key Laboratory of Optical Astronomy, National Astronomical Observatories, Chinese Academy of Sciences, Beijing 100101, People's Republic of China; xtwfn@bao.ac.cn}
\affiliation{School of Astronomy and Space Science, University of Chinese Academy of Sciences, Beijing 100049, China}

\correspondingauthor{Wei Zhang}
\email{xtwfn@bao.ac.cn}

\begin{abstract} 

Using data from the LAMOST Medium-Resolution Spectroscopic Survey of Nebulae, we create a sample of 17,821 diffuse ionized gas (DIG) spectra in the anti-center region of the Milky Way, by excluding fibers in the directions of H II regions and supernova remnants. We then analyze the radial and vertical distributions of three line ratios ([N II]/H$\alpha$, [S II]/H$\alpha$, and [S II]/[N II]), as well as the oxygen abundance.

[N II]/H$\alpha$ and [S II]/H$\alpha$ do not exhibit a consistent, monotonic decrease with increasing Galactocentric distance (R$_{gal}$). Instead, they show enhancement within the interarm region, positioned between the Local Arm and the Perseus Arm. [S II]/[N II] has a radial gradient of 0.1415 $\pm$ 0.0646 kpc$^{-1}$ for the inner disk (8.34 $ < R_{gal} < $ 9.65 kpc), and remains nearly flat for the outer disk ($R_{gal} > $ 9.65 kpc).  In the vertical direction, [N II]/H$\alpha$, [S II]/H$\alpha$, and [S II]/[N II] increase with increasing Galactic disk height ($|z|$) in both southern and northern disks.

Based on the N2S2H$\alpha$ method, which combines [S II]/[N II] and [N II]/H$\alpha$, we estimate the oxygen abundance. The oxygen abundance exhibits a consistent radial gradient with R$_{gal}$, featuring a slope of -0.0559 $\pm$ 0.0209 dex kpc$^{-1}$ for the inner disk  and a similar slope of -0.0429 $\pm$ 0.0599 dex kpc$^{-1}$ for the outer disk. A single linear fitting to the entire disk yields a slope of -0.0317 $\pm$ 0.0124 dex kpc$^{-1}$. In the vertical direction, the oxygen abundance decreases with  increasing $|z|$ in both southern and northern disks.

\end{abstract}


\section{Introduction}

The interstellar medium (ISM) of the Milky Way consists of both interstellar gas and dust. Interstellar gas plays an important role in the form and death of stars as well as the evolutionary cycle of the interstellar medium. The properties of interstellar gas are affected by the radiation and matter ejected from stars, which in turn influence the characteristics of new generation stars. Interstellar gas can be further categorized into cold neutral gas comprising atomic and molecular gas, and ionized gas \citep{Sparke_Linda_2007}. Ionized gas encompasses several components: the extremely hot ionized galactic halo, HII regions, supernova remnants (SNRs), planetary nebulae (PNe), and diffused ionized gas (DIG) \citep{Draine-11}. DIG was first discovered from the free-free absorption when surveying the Galactic synchrotron background at radio wavelengths \citep{HoyleF_1963}. Later, faint emission lines originated from DIG were observed at optical wavelengths confirming its existence \citep{Reynolds_1973}. It has been estimated that DIG constitutes approximately 20$\%$ of the overall gaseous content while representing up to 90$\%$ among all ionized gases in Milky Way, thus established as a primary source for such ions within our galaxy \citep{Reynolds_1991, Haffner_2009}.

Some surveys have been designed to explore the Galactic DIG. For example, the Wisconsin H-Alpha Mapper (WHAM) Sky Survey in optical band \citep{Haffner_2003}, and the Green Bank Telescope Diffuse Ionized Gas Survey (GDIGS) in radio band \citep{Anderson_2021}. Additionally, data on piggyback recombination lines (RRLs)  from  the ongoing Galactic Plane Pulsar Snapshot (GPPS) survey offer valuable insights for analyzing Galactic DIG \citep{HouLigang_2022}. Various emission lines have been used to investigate the distribution and characteristics of DIG in different bands. In the optical bands, [N II], [S II], [O II], [O III], O I, He I and H$\alpha$ lines are employed \citep{Reynolds_1998, Haffner_1999ApJ, Hausen_2002, Madsen_2006ApJ}. While in radio bands, RRLs of hydrogen, helium, and carbon serve as crucial probes \citep{Luisi_2017, Luisi_2019, Anderson_2021}.
The intensity of diffuse Spitzer GLIMPSE 8.0 $\mu$m emission is found to be correlated with the intensity of the RRL emission from the DIG \citep{Luisi_2017}. The mid-infrared fine structure lines of [Ne II], [S III], and possibly [S IV] have be detected by the James Webb Space Telescope (JWST) \citep{Kulkarni_2024}. DIG can also be detected in the fine structure far-infrared lines (eg., [C II] and [N II]) in absorption  \citep{Persson_2014, Gerin_2015}, or in emission \citep{Velusamy_2012, Velusamy_2015, Langer_2017}.

WHAM data has provided valuable insights into the physical properties and spatial distribution of Galactic DIG. Notably, the detection of the [O I]$\lambda$6300 emission line within DIG regions near the Galactic plane indicates a predominance of ionized hydrogen, as inferred from the low [O I]/H$\alpha$ ratio \citep{Reynolds_1998, Hausen_2002}. Furthermore, investigations have revealed that the [S II]/H$\alpha$ and [N II]/H$\alpha$ ratios exhibit a tendency to increase with decreasing H$\alpha$ intensity, a trend that is accentuated with increasing vertical distance from the Galactic plane ($|z|$) \citep{Reynolds_1999}. Comparisons with classical HII regions have highlighted distinct differences in the ionization characteristics of DIG. Specifically, DIG displays higher [N II]/H$\alpha$ and [S II]/H$\alpha$ ratios, coupled with lower [O III]/H$\alpha$ and He I/H$\alpha$ ratios. These findings suggest that DIG regions tend to have higher temperatures, exist in a lower ionization state, and are likely ionized by a softer spectrum of radiation compared to classical H II regions \citep{Madsen_2006ApJ}.

WHAM has conducted a comprehensive all-sky survey of H$\alpha$ emission from the Milky Way, however, other crucial emission lines necessary for inferring the properties of DIG are only observed in selected regions of the sky. Consequently, the continued distribution of DIG along the Galactic disk remains an open question. It is widely acknowledged that the metallicity of HII regions exhibits a negative radial gradient on the Galactic plane \citep[e.g.][]{Balser_2011,Esteban_2017, Mendez-Delgado_2022}. As there are various possible ionize sources for DIG, such as leaking photons from HII regions \citep{Haffner_2009}, hot low-mass evolved stars (HOLMES) \citep{Flores-Fajardo_2011, Yan_2012} and shocks \citep{Collins_2001}, hereafter, the distribution of DIG does not necessarily associated with HII regions \citep{Luisi_2017}.  Recent observations of M83 by MUSE have found that the fraction of H$\alpha$ luminosity originating from the DIG varies with galactic radius, and peaks in the interarm regions  \citep{Della_Bruna_2022}. Consequently, in this study, we aim to investigate the distribution patterns of various parameters, such as [N II]/H$\alpha$, [S II]/H$\alpha$ and [S II]/[N II] ratios, as well as oxygen abundance, along the Galactic disk within the DIG.

The contamination of DIG can significantly impact the characteristics of HII regions. This contamination can lead to misclassification of positions in the Baldwin–Phillips–Terlevich (BPT) diagram \citep{ZhangKai_2017}, as well as overestimation of metallicity measurements \citep{Sanders_2017}. Conversely, this contamination also affects the analysis of DIG properties. According to the Galactic HII region catalog \citep{Anderson_2014}, the majority of HII regions have angular sizes smaller than several arcminutes. Therefore, achieving spatial resolutions at least at the arcminute level is necessary to effectively distinguish HII regions from DIG. The Large Area Multi-Object fiber
Spectroscopic Telescope (LAMOST)  has started the medium-resolution survey (LAMOST-MRS) in October 2018 \citep{Liu-20, Wang-96, Su-Cui-04, Cui-12, Zhao-12, Luo-15}. The ongoing project of LAMOST-MRS of Nebulae (LAMOST MRS-N)  provides an excellent opportunity to study Galactic DIG, considering its high spatial resolution ($\sim 3\arcmin$)\footnote{In this paper, spatial resolution is defined by the median fiber-to-fiber distance. Each fiber has a size of 3$\farcs$3, with significant spacing between them, resulting in a median distance of 3$\arcmin$ between adjacent.}, and wide sky coverage ($\sim$ 1700 $deg^2$) \citep{WuChaojian_2021, RenJuanjuan_2021, ZhangWei_2021, WuChaojian_2022}. Based on this survey, we can construct a large sample of Galactic DIG to investigate the distribution of DIG on the disk, including radial and vertical gradients of DIG parameters along the Galactic disk.

The paper is structured as follows. The criteria for selecting Galactic DIG are described in Section 2. The methods for deriving kinematic distance and oxygen abundance are introduced in Section 3. In Section 4, we investigate the distributions of line ratios, and oxygen abundance along and perpendicular to the Galactic disk. We discuss the results in Section 5 and present the conclusion in Section 6.

 \begin{figure*}[htbp]
	\centering
	\includegraphics[scale=0.4]{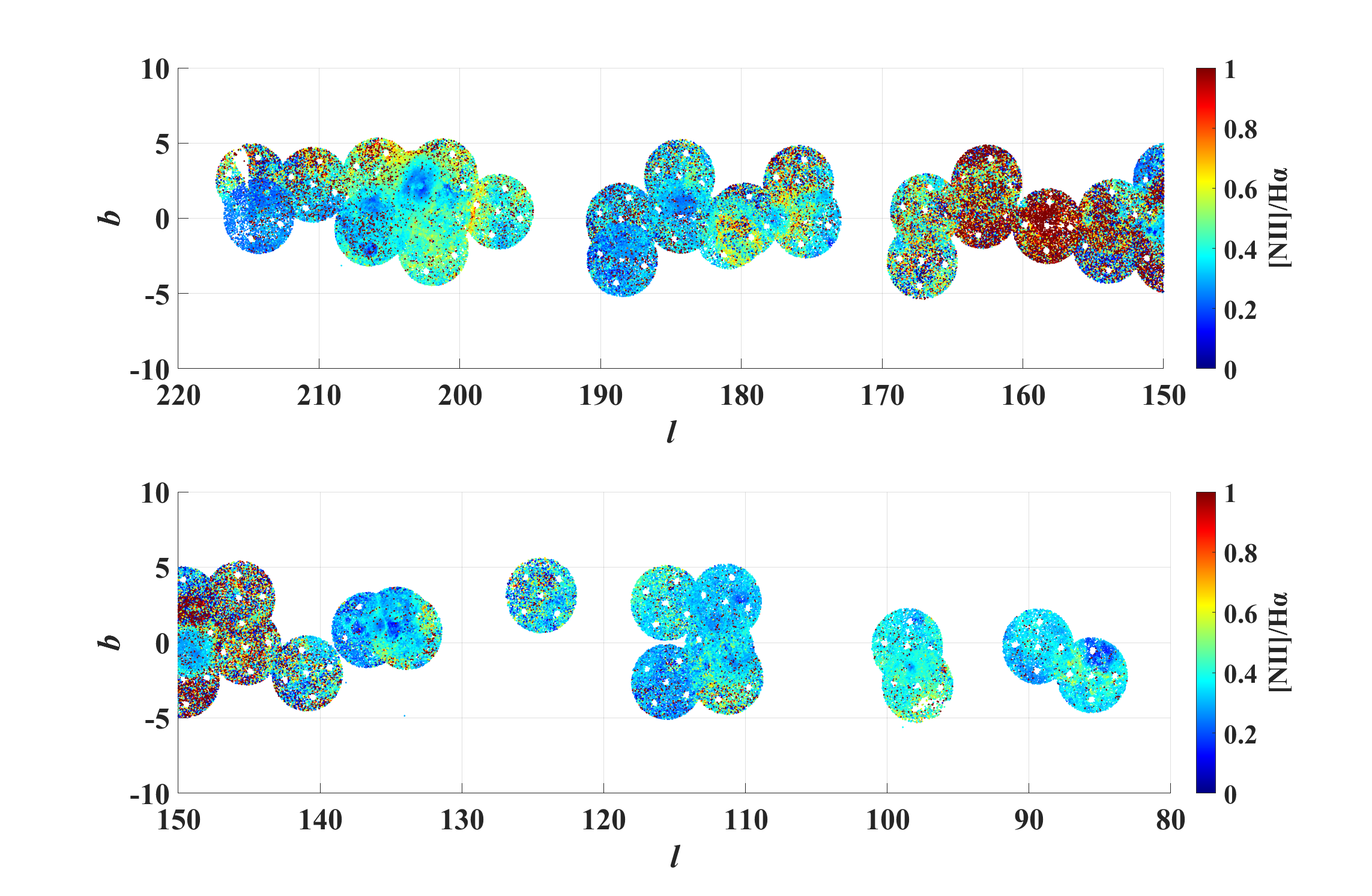}
	\includegraphics[scale=0.4]{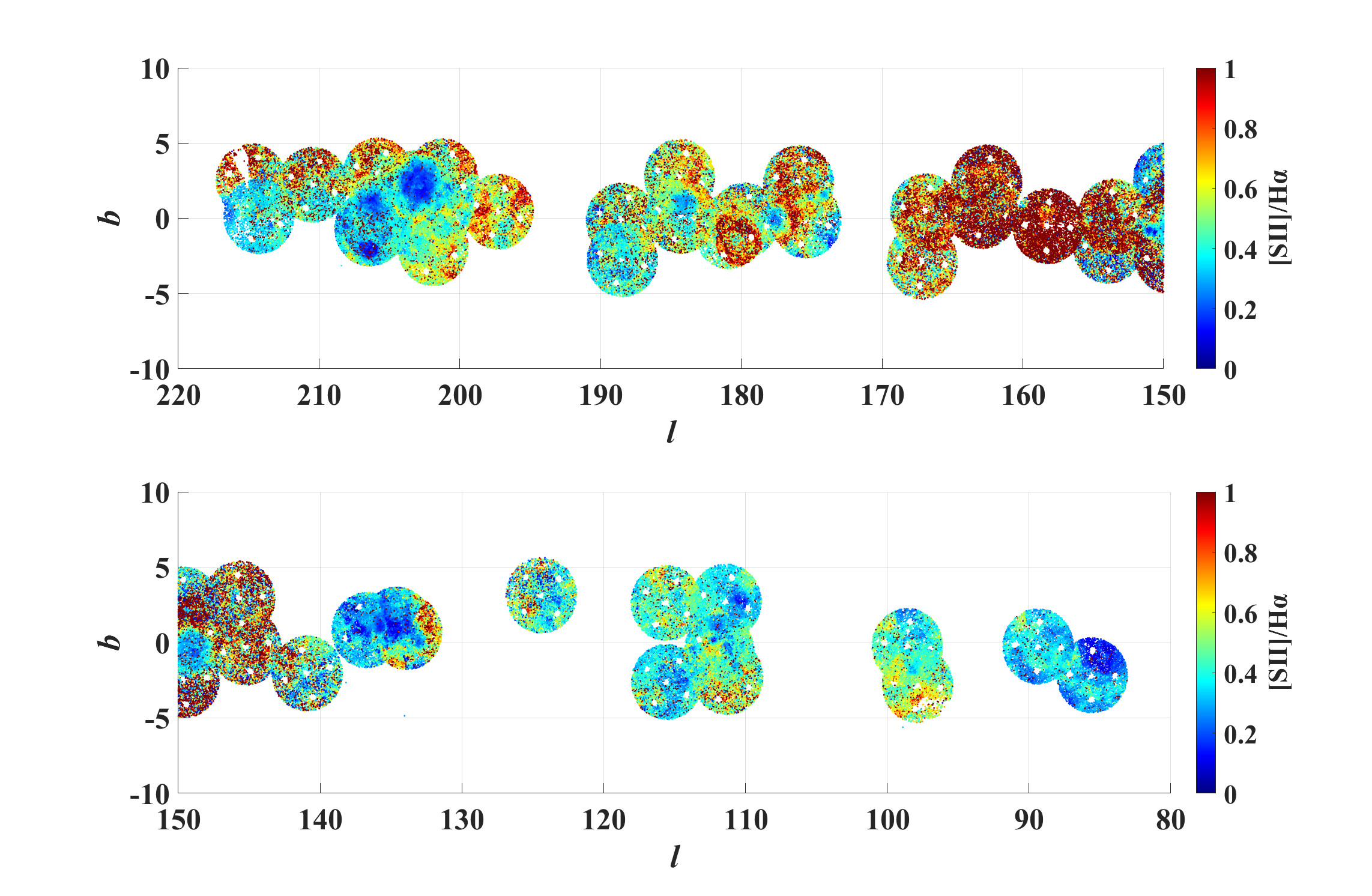}
	\caption{
	Footprints of LAMOST MRS-N up to the end of 2021, colored by [N II]/H$\alpha$ ratios in the top two panels and [S II]/H$\alpha$ ratios in the bottom two panels.
	}
	\label{fig:ObservedRange}
\end{figure*}

\section{Sample construction}

The primary goal of LAMOST MRS-N is to survey ionized gas and bright stars in the Galactic
disk ($|b| \le 5^{\circ}$), within the Galactic longitude range of 80$^{\circ} < l <
220^{\circ}$. The optical spectrum spans from 4950 to 5350 \AA~  in the blue
channel and 6300 to 6800 \AA~ in the red channel, featuring a resolution power
of $R = \lambda/\Delta\lambda \sim$ 7500, as outlined by
\citep{WuChaojian_2021}.  For detailed information
regarding the recalibration of wavelengths using skylines, readers are referred
to the study by \cite{RenJuanjuan_2021}. Additionally, the method of subtracting
geocoronal H$_{\alpha}$ emission (H$_{\alpha,sky}$) by exploiting the
correlation between the ratio of H$_{\alpha,sky}$ and OH$\lambda$6554 line
intensities and solar altitude is described in the work of
\citet{ZhangWei_2021}. The throughput pipeline for data reduction and the 
resulting data products are summarized in the work by \citet{WuChaojian_2022}.

We specifically focuses on four emission lines in the
red channel, namely [N II]$\lambda$6584, H$\alpha$, [S
II]$\lambda\lambda$6717,6731. Due to the absence of flux calibration for LAMOST
MRS-N, three key ratios will be utilized in subsequent analyses: Flux([N
II]$\lambda$6584)/Flux(H$\alpha$) (hereafter [N II]/H$\alpha$), Flux([S
II]$\lambda\lambda$6717,6731)/Flux(H$\alpha$) (hereafter [S II]/H$\alpha$), and
Flux([S II]$\lambda\lambda$6717,6731)/Flux([N II]$\lambda$6584) (hereafter [S
II]/[N II]). 

The sky coverage of fields obtained by LAMOST MRS-N to the end of 
2021 have been shown in Figure~\ref{fig:ObservedRange}, with color
representation based on the [N II]/H$\alpha$ line ratio. During observations, some regions are overlapped, resulting in these areas being observed multiple times. As these overlapping regions are over-sampled compared to areas observed only once, it is necessary to resample the data to a uniform density.  Given a median spatial resolution of 3\arcmin, and considering that most fibers have distances less than 5\arcmin, we resample the data into 5\arcmin$\times$5\arcmin~bins. For each 2D bin, if only one fiber is found, we retain this fiber in the new dataset and assign the center of the bin as its new coordinate. In cases where more than one fiber are found within a bin, we retain the fiber with the highest Signal-to-Noise (S/N) ratio of H$\alpha$ emission and assign the center of the bin as its new coordinate.

\begin{figure*}[htbp]
	\centering
	\includegraphics[scale=0.4]{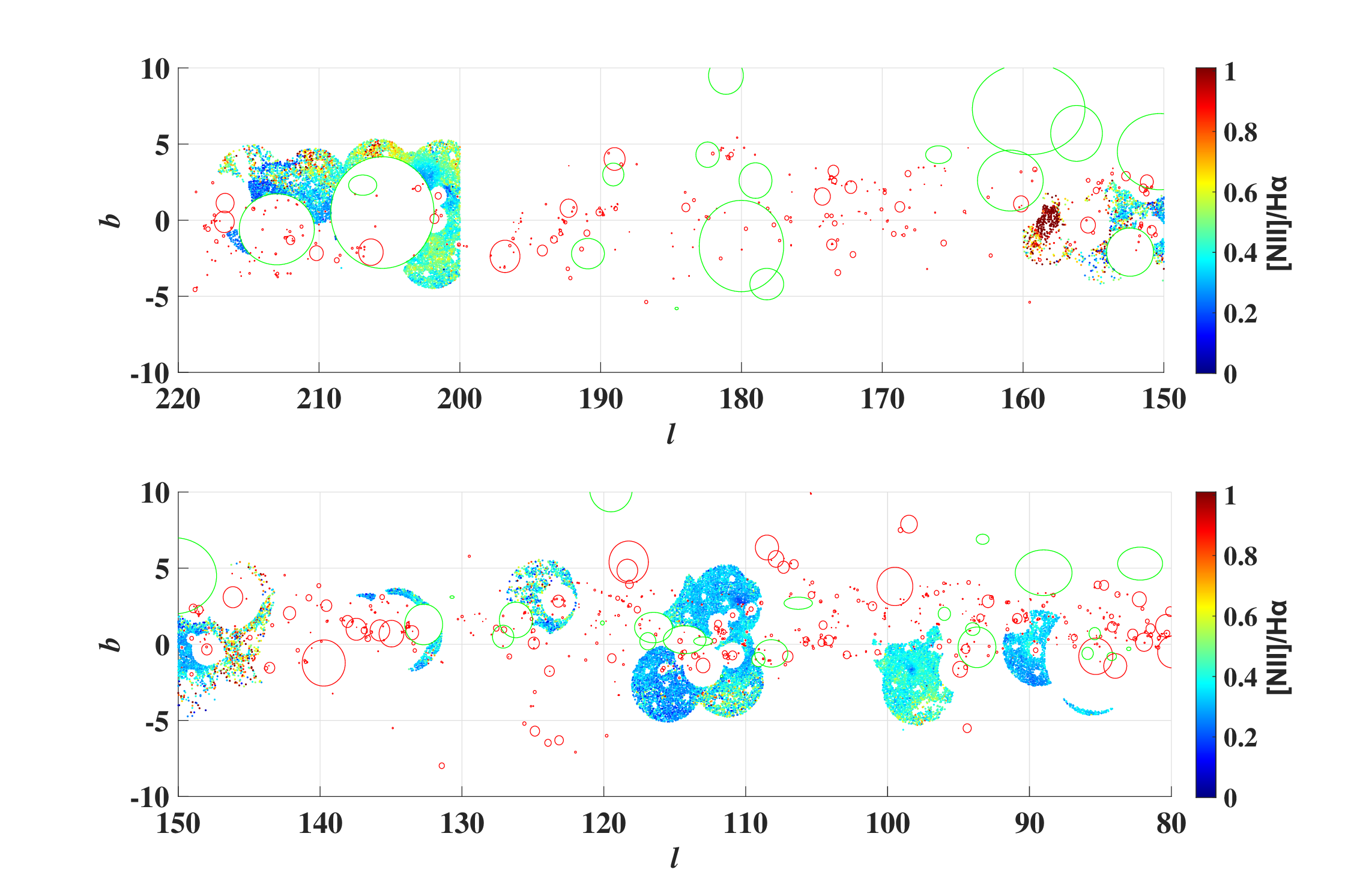}
	\includegraphics[scale=0.4]{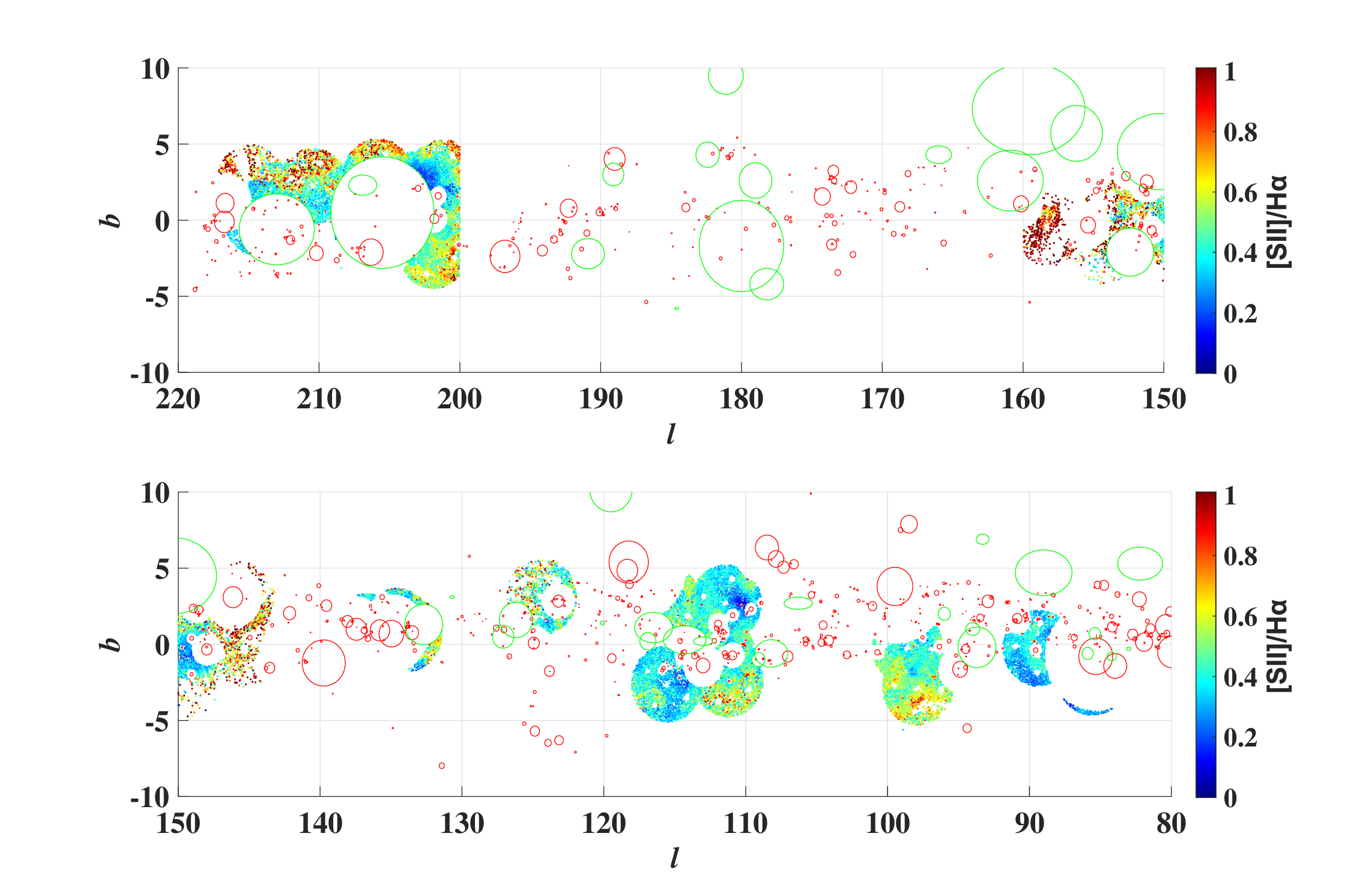}
	\caption{Spatial distribution of the Galactic DIG sample (Sample-I). Red circles represent the HII regions sourced from the WISE HII catalog by \cite{Anderson_2014}, while the green circles correspond to the SNRs listed in the catalog by \citet{GreenD_2019}. Upper two panels: colored by [N II]/H$\alpha$ ratios; bottom two panels: colored by [S II]/H$\alpha$ ratios.}
	\label{fig:3Rsamples}
\end{figure*}

\begin{figure*}[htbp]
	\centering
	\includegraphics[scale=0.4]{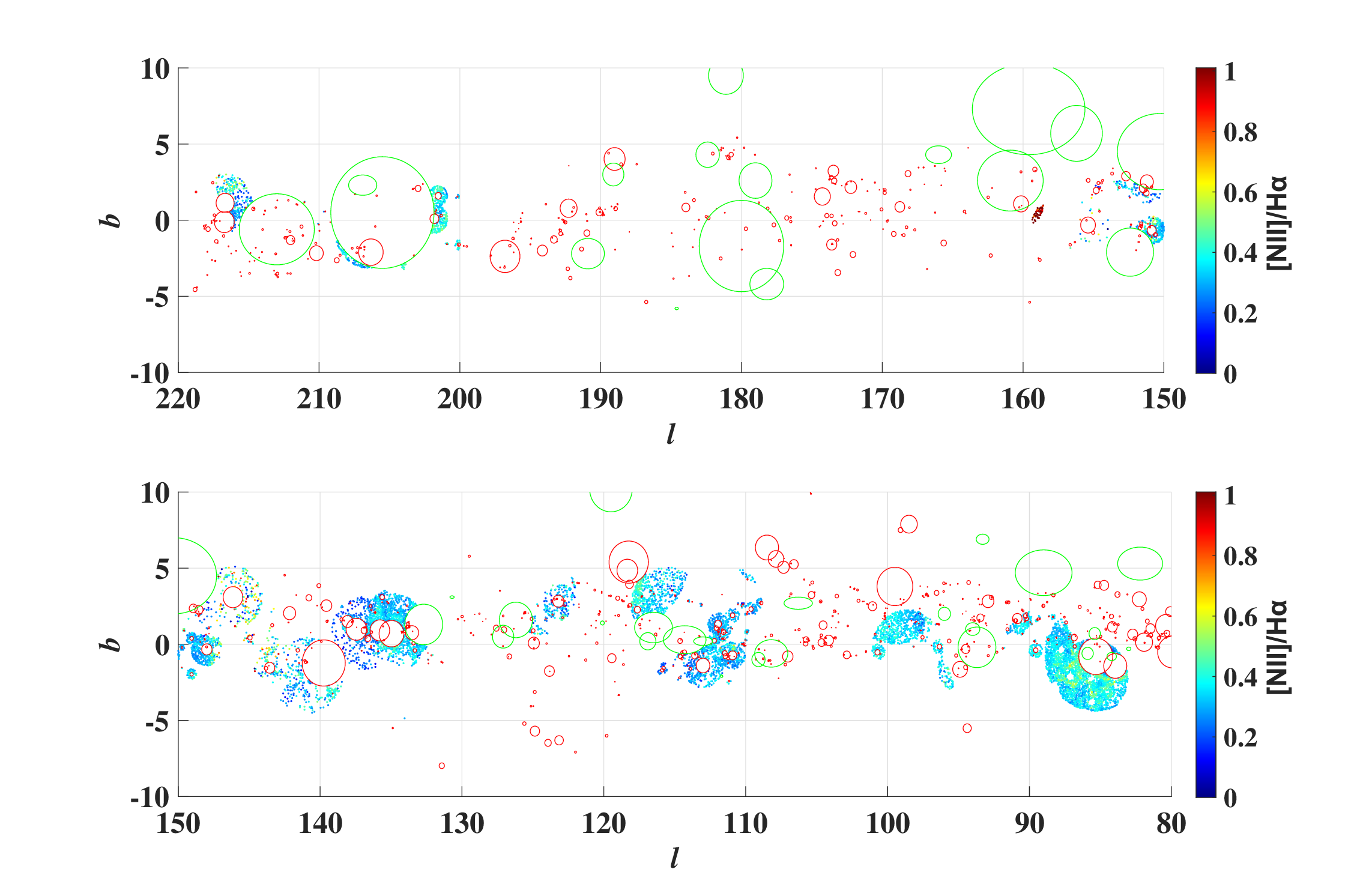}
	\includegraphics[scale=0.4]{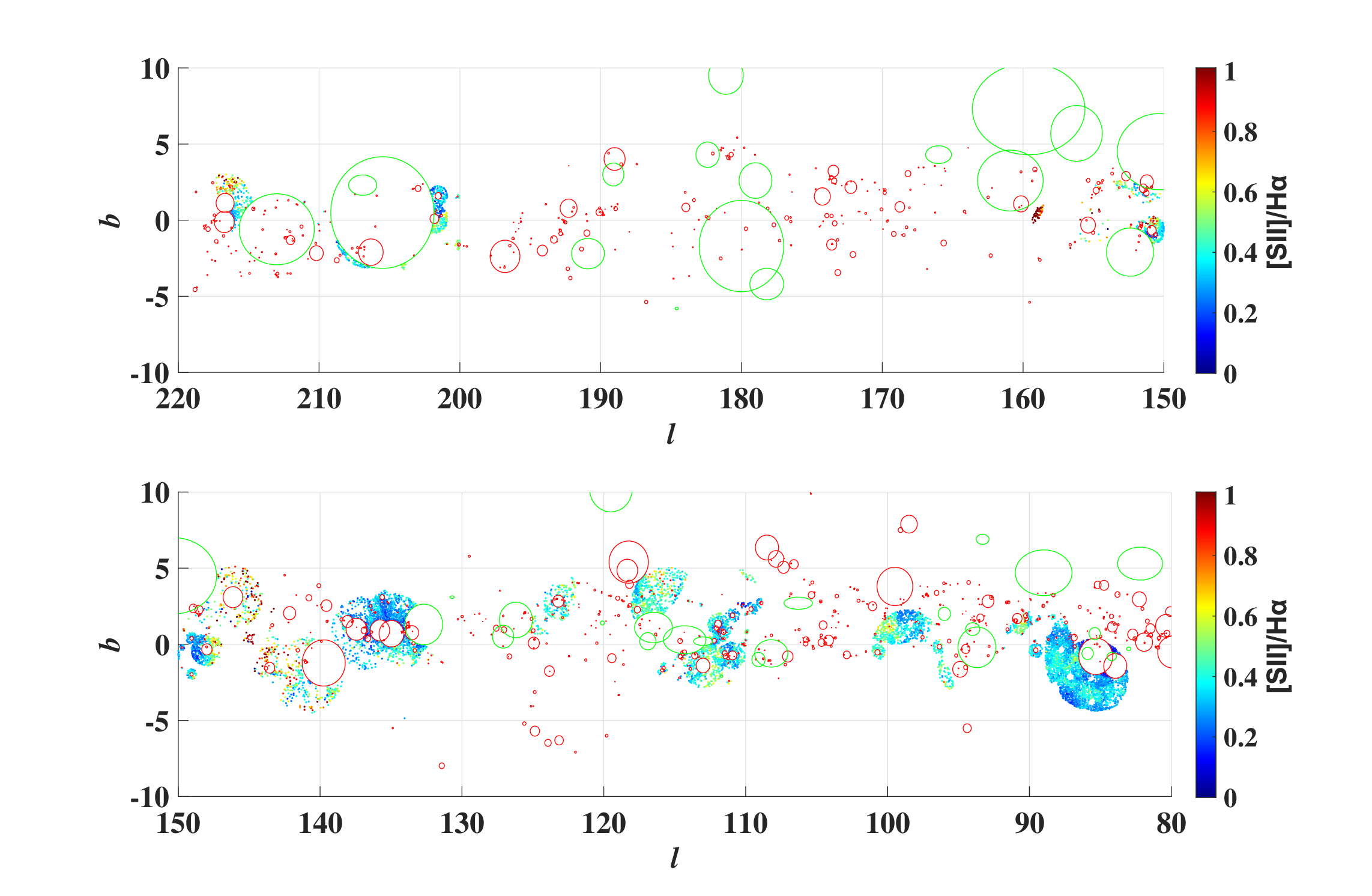}
	\caption{Same as Figure ~\ref{fig:3Rsamples}, but for the transition sample (Sample-II). Upper two panels: colored by [N II]/H$\alpha$ ratios; bottom two panels: colored by [S II]/H$\alpha$ ratios.}
	\label{fig:1_3Rsamples}
\end{figure*}	
	
\begin{figure*}[htbp]
	\centering	
	\includegraphics[scale=0.4]{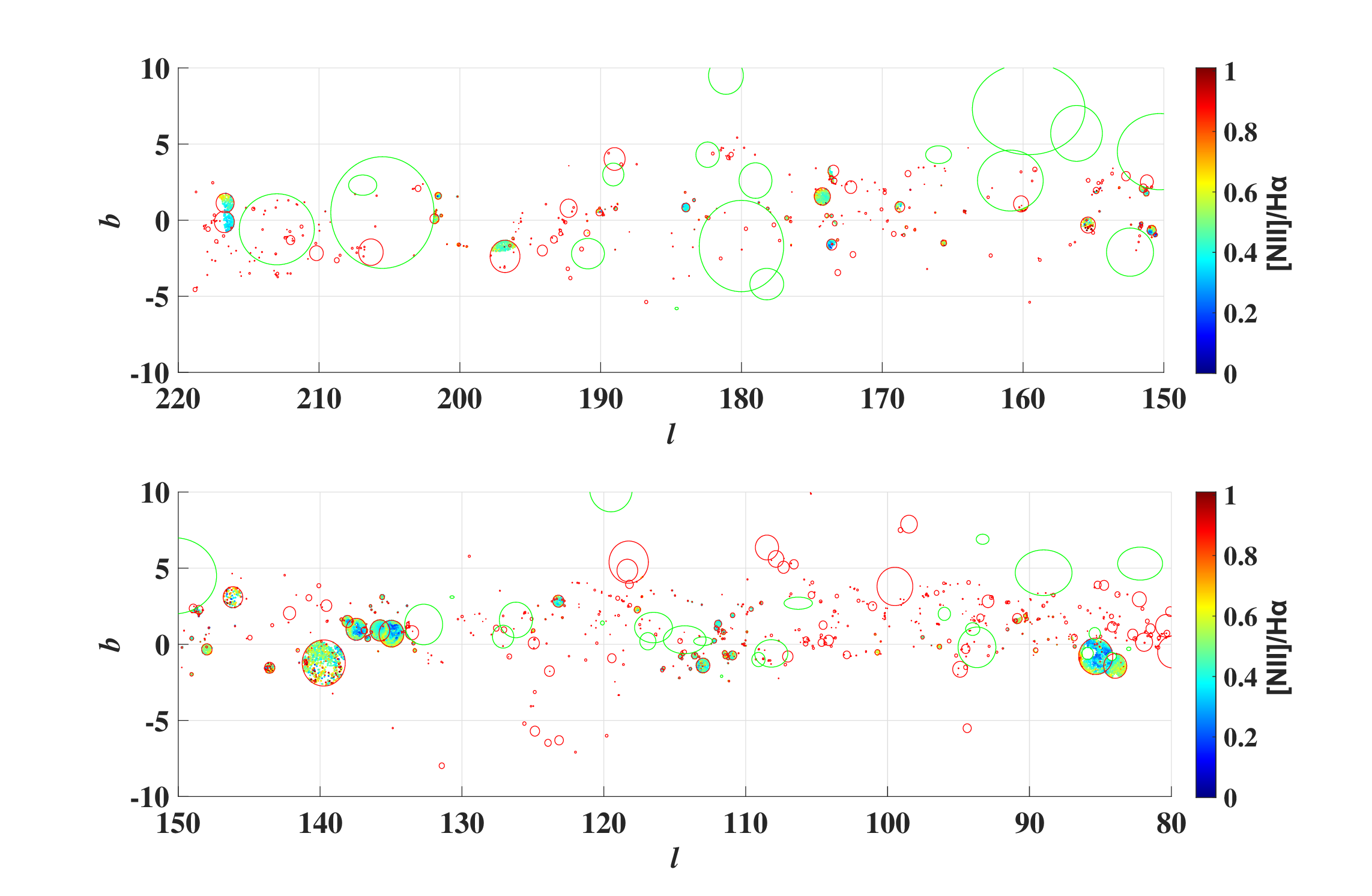}
	\includegraphics[scale=0.4]{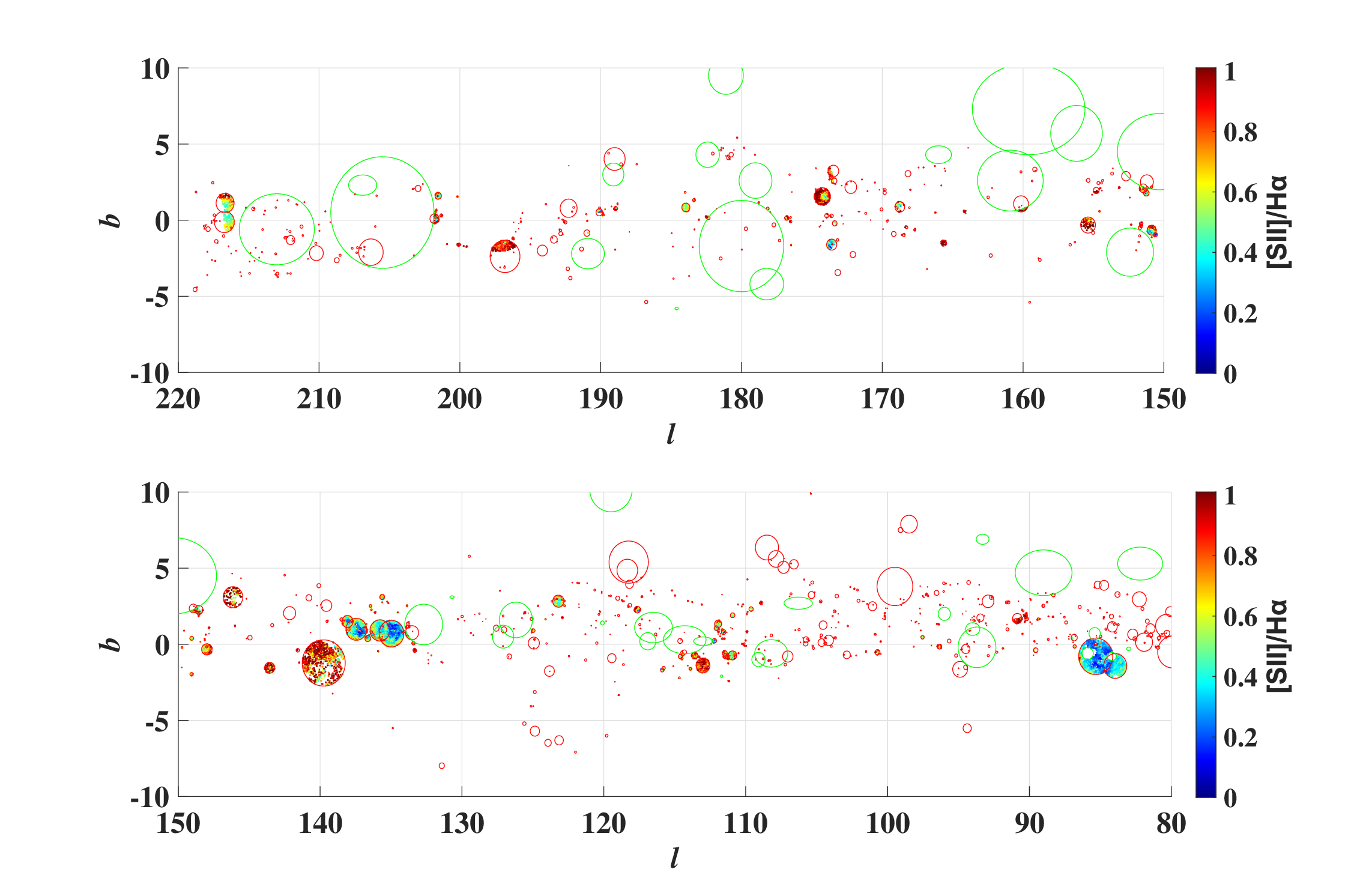}
	\caption{Same as Figure ~\ref{fig:3Rsamples}, but for  HII regions. Upper two panels: colored by [N II]/H$\alpha$ ratios; bottom two panels: colored by [S II]/H$\alpha$ ratios.}
	\label{fig:1Rsamples}
\end{figure*}

\begin{figure}[htbp]
	\centering
	\includegraphics[scale=0.18]{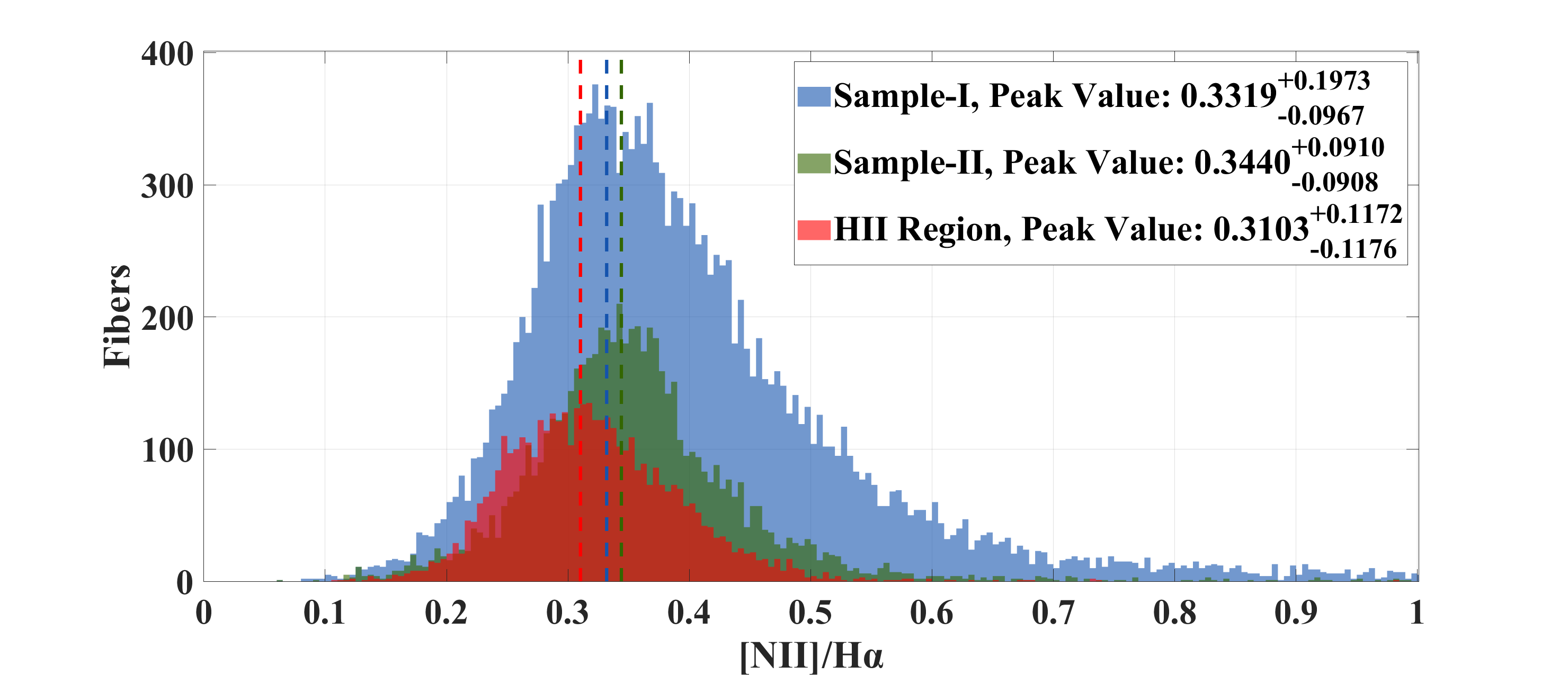}
	\includegraphics[scale=0.18]{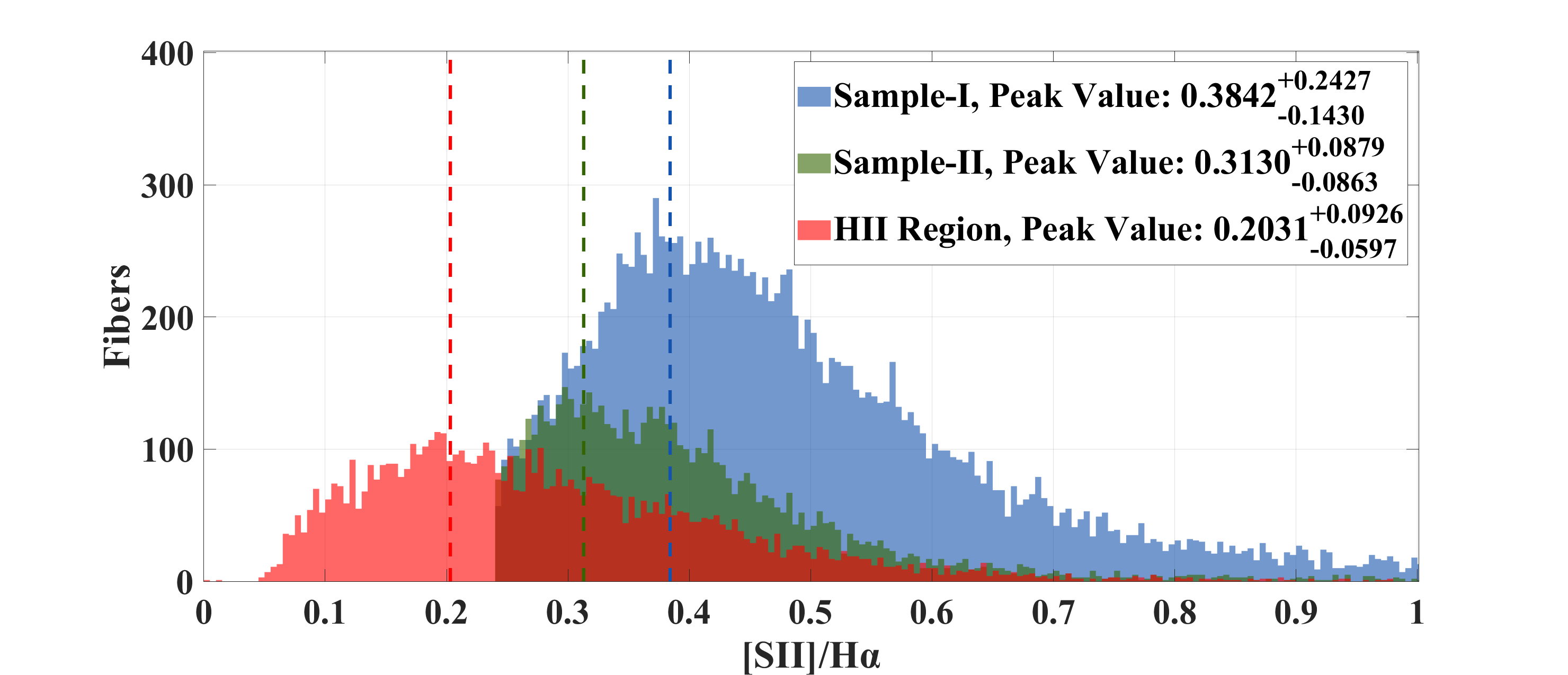}
	\caption{Histogram distribution of [N II]/H$\alpha$ and [S II]/H$\alpha$ line ratios. The panels represent Sample-I, Sample-II, and HII regions with blue, green and red bars, respectively. Dashed line indicates the distribution's peak. The regions to the left and right of this peak are individually fitted with Gaussian profiles. The 1$\sigma$ values of the Gaussians are annotated in the upper-right corner of each panel.}
	\label{fig:Statistical_distributions}
\end{figure}

Various methods have been employed to distinguish between HII regions and DIG. These approaches are primarily rely on the H$\alpha$ surface brightness \citep{Kaplan_2016, ZhangKai_2017}, the H$\alpha$ line equivalent width \citep{Belfiore_2016, Kumari_2019}, or the [N II]/H$\alpha$ and [S II]/H$\alpha$ line ratios (\citet{Kumari_2019}, empolying the classical BPT diagrams \citep{Baldwin_1981, Kauffmann_2003, Kewley_2006}. If the spaitial resolution is sufficiently high, an effective strategy for isolating DIG involves the identification of individual H II regions and subsequent subtraction \citep{Belfiore_2022}.

In this work, we cross-match the LAMOST MRS-N database with the current largest  
Galactic HII cataloge selected from WISE MIR emission \citep{Anderson_2014,
Wright_Edward_2010}, and eliminate fibers located within three times the radius of
the HII regions ($\rm r_{WISE}$) to minimize contamination from corresponding HII regions.  
We also exclude the fibers located within the SNR regions defined by $\frac{(x - x_{0})^{2}}{a^{2}} + \frac{(y - y_{0})^{2}}{b^{2}} \le 1$, 
where $a$ and $b$ are the semi-major and semi-minor axes of the ellipse and $(x_{0}, y_{0})$ is the central position of the SNR, according to the Galactic SNRs catalog by \citet{GreenD_2019}.

Additional criteria have also been considered as follows:  
\begin{itemize}
\item [(1)] All four emission lines are required to have a signal-to-noise ratio (S/N)  greater than 3.

\item [(2)] $\rm |V_{[N\,II]} - V_{H\alpha}| < 20\, km\,s^{-1}$ \\   
 $\rm |V_{[S\,II]} - V_{H\alpha}| < 20\, km\,s^{-1}$ \\     
 $\rm |V_{[N\,II]} - V_{[S\,II]}| < 20\,km\,s^{-1}$ \\
 This criterion is intended to eliminate fibers that may have a large systematic error in measuring radial velocity.
 
\item [(3)] $\rm 0.24 < [S\,II]/H\alpha$ \\ 
The fibers located in the direction of HII regions exhibit a peak at $\sim$ 0.20 for [S II]/H$\alpha$ ratio, and the distributions of DIG and HII regions intersect at $\sim$ 0.24 (Zhao et al., in preparation, see also Figure~\ref{fig:Statistical_distributions}). Therefore, the range 0.24 $<$ [S II]/H$\alpha $ is employed to further exclude potential HII regions from the analysis.

\item [(4)] $80^{\circ} < l < 160^{\circ}$ and $200{^\circ} < l < 220^{\circ}$\\                                     
The sources in the range of $160 \le l \le 200 $ have been removed due to the uncertainties in kinematic distance measurement being too large in this direction.
\end{itemize}

Finally, we construct a large sample of Galactic DIG, comprising 17,821 fibers, hereafter referred to as Sample-I. The spatial distribution of this sample are illustrated in Figure~\ref{fig:3Rsamples}, colored by the [N II]/H$\alpha$ and the [S II]/H$\alpha$ line ratio. Red circles denote the region defined by the WISE HII catalog, while green circles represent the regions of SNRs as defined by \citet{GreenD_2019}.

In order to compare the properties of DIG with the gas adjacent to HII regions and infer the effects of the HII regions, we construct another sample. This sample is selected using similar criteria, with the additional requirement that the fibers are located within 1-3 times the HII radii, and also remove fibers with [S II]/H$\alpha$ $<$ 0.24. This resulting sample consists of 6,925 fibers and is denoted as Sample-II. Sample-II can be considered as the transitional regions between HII regions and DIG regions. By applying criteria (1) and (2) and ensuring that at least one fiber is located within the WISE HII radius region, we constructed a large sample of $\sim$ 300 HII regions as targets. We retained sources within the range of $160 \leq l \leq 200 $, as we opted to utilize combined distance measures such as parallax distance of ionizing OB central stars, 3D extinction distance, velocity dispersion versus size relation, and kinematic distance when uncertainty is small enough. Since the S/N of the spectra for HII regions is relatively higher than that of DIG, and to retain more spectra for the HII regions, we resample the HII sample into 3\arcmin$\times$3\arcmin~bins. As a result, the HII regions consist of 5,420 individual fibers.
Details of this sample will be presented in Zhao et al. (in preparation). The spatial distribution of  sample-II and HII regions are shown in Figure~\ref{fig:1_3Rsamples}.

The histogram distributions of the line ratios of [N II]/H$\alpha$ and [S II]/H$\alpha$ of Sample-I are showed in  Figure~\ref{fig:Statistical_distributions} as blue bars. To facilitate a comparison between DIG, HII regions and the transitional regions between them, the histogram distributions of these two line ratios of Sample-II and HII regions are presented in the same panels as green and red bars. 

The analysis reveals that the DIG sample have exhibits a [N II]/H$\alpha$ flux ratio with a peak at 0.33, a little higher than that for HII regions which peaks at 0.31. For the [S II]/H$\alpha$ line ratio of DIG displays a peak at 0.38, is significantly higher than HII regions of 0.20. This result is consistent with the well-established finding that DIG exhibits enhanced [N II]/H$\alpha$ and [S II]/H$\alpha$ ratios compared to classical HII regions \citep[e.g.,][]{Madsen_2006ApJ}, suggesting that DIG is warmer than HII regions \citep[e.g.,][]{Haffner_2009} and that the DIG spectrum is dominated by low ionization states \citep{Zurita-2000}.
Besides, compared to Sample-II and HII regions, Sample-I do not show a much wider distribution range of the line ratios of [S II]/H$\alpha$ and [N II]/H$\alpha$.

\begin{figure*}[htbp]
	\centering

	\includegraphics[scale=0.257]{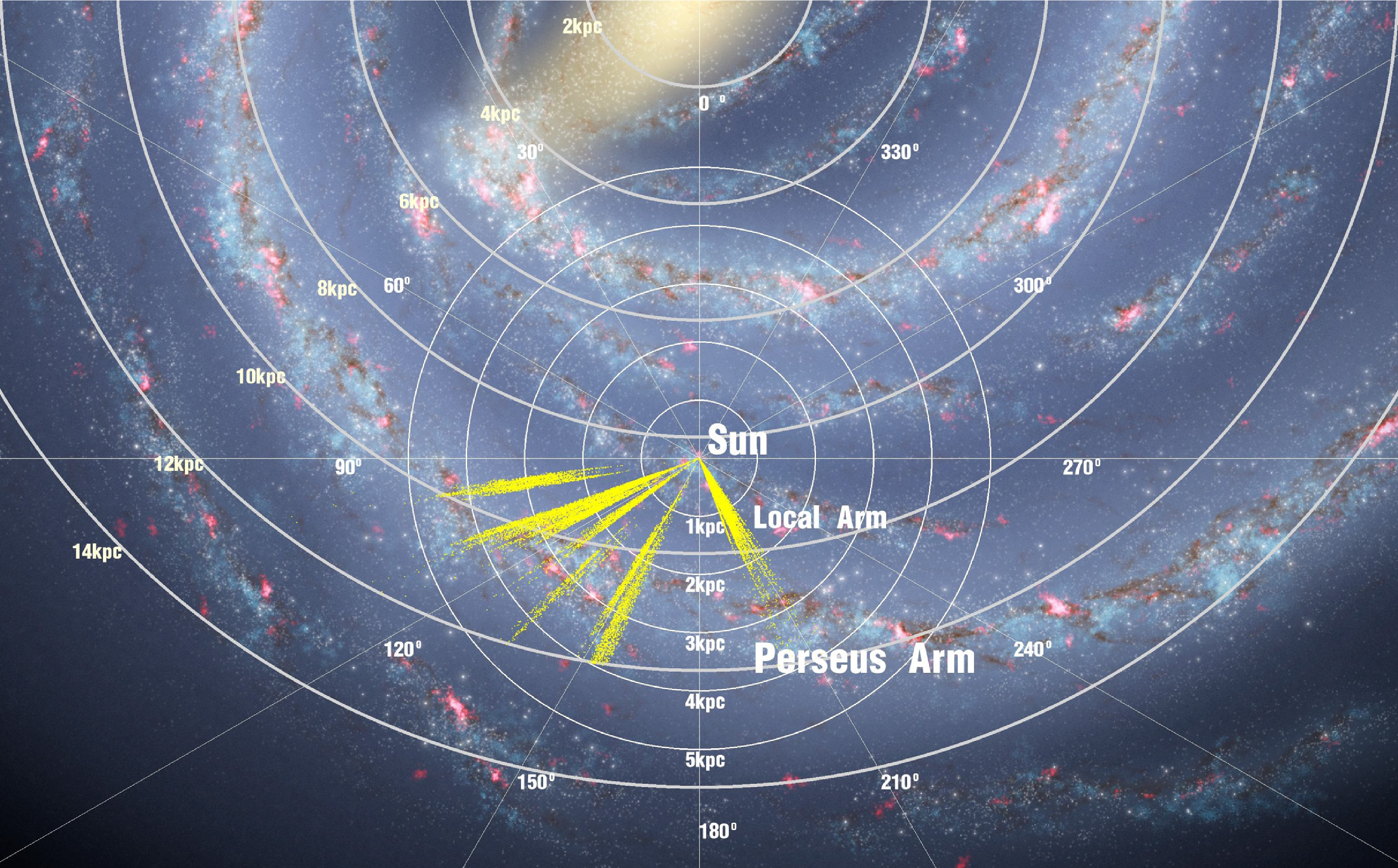}
	
	\caption{Plan view of the disk DIG from the north Galactic pole, display the projection of the sample points on the galactic disk. The Galactic center is taken as the origin point, and the direction from the Sun to the Galactic center is taken as the polar coordinate starting direction, and radial unit is kpc. We mark the location of Solar system at the Galactocentric distance of 8.34 kpc. Background image credit: NASA/JPL-Caltech/ESO/R. Hurt 
	}
	\label{fig:DiskDistribution}
\end{figure*}

\section{Parameter Calculation}
\subsection{Kinematic Distance}

As the ionization source is not always specified -- often being a combination of contributions from HII regions, HOLMES, SNRs, and others -- it is challenging to identify the ionization source and use its distance as the distance to Galactic DIG. Simultaneously, DIG tends to be located in lower-density environments, making it impossible to derive the distance based on the 3D reddening map. Therefore, the only viable method, and perhaps the most plausible one, is the kinematic distance method. Although DIG in LAMOST MRS-N does not suffer from the kinematic distance ambiguity (KDA), the uncertainty of such distances can be as high as 50\% for sources within 20 degrees of the Galactic anti-center \citep{Anderson_2014}. This is why we have removed the sources in the range of $160^\circ \le l \le 200^\circ$ from our sample. The local standard of rest (LSR) velocity (VLSR) has been transferred from heliocentric velocity by adopting the motion of the Sun in the LSR, which is moving at a speed of about 20 $\rm km\,s^{-1}$ towards (RA = 18h03m50.29s, DEC = +30\arcdeg00\arcmin16.8\arcsec) at epoch 2000. The kinematic distances are derived using the Python package downloaded from https://github.com/tvwenger/kd \citep{Wenger_2018}. During the calculation, the rotation curve by \citet{Reid_MJ_2014} is adopted. The Galactocentric distance (R$_{gal}$) is calculated by adopting R$_\odot$ = 8.34 kpc. The plan view of the Sample-I is shown in Figure~\ref{fig:DiskDistribution}.

\subsection{Oxygen Abundance}

The wavelength range covered by the LAMOST MRS-N data enables us to use three indirect metallicity calibrators, N2  \citep{Pettini_2004}, S2 \citep{Curti_mirko_2020} and N2S2H$\alpha$ \citep{Dopita_2016}. However, N2 method is not applicable in the DIG region, as the N2–metallicity calibration saturates and flattens at the BPT boundary between H II regions and low ionized regions (LIERs) \citep{Curti_mirko_2020, Kumari_2019}. In this work, we prefer to use [N II]/H$\alpha$ and [S II]/H$\alpha$ line ratios to investigate their variations across the Galactic disk rather than employing them to estimate the metallicity. Given N2S2H$\alpha$ method is insensitivity to changes in the ionization parameter, owing to the inclusion of the [N II]/[S II] emission line ratio \citep{Poetrodjojo_2019}, we employ the N2S2H$\alpha$ method to estimate the oxygen abundance as follows:

 \begin{equation}
\begin{split}
\rm
12 + log (O/H) = 8.77 + log ([N\,II]/[S\,II]) \\ 
\rm
 + 0.264 \times log ([N\,II]/H\alpha)
	\label{N2S2_method} 
\end{split}
\end{equation}

\section{Radial and vertical distributions of line ratios, and oxygen abundance}

\subsection{Radial distribution}

\begin{figure}[htbp]
	\centering
	
    \includegraphics[scale=0.3]{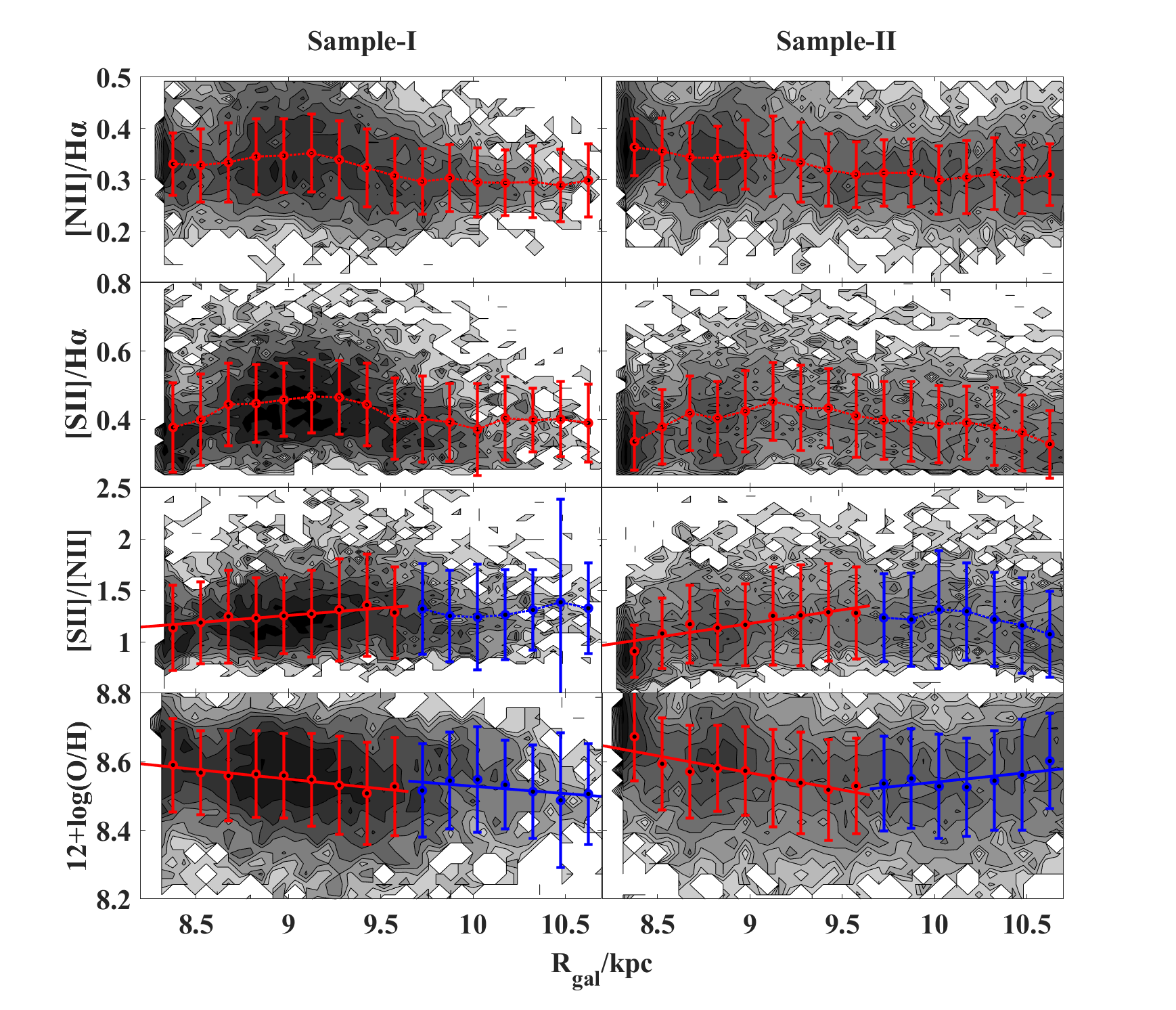}
	\caption{Radial distributions of [N II]/H$\alpha$, [S II]/H$\alpha$, [S II]/[N II] line ratios, and oxygen abundance as a function of R$_{gal}$. The data are organized into 16 bins based on R$_{gal}$, and a dashed line connects the median value of each bin. The results of Sample-I and Sample-II are illustrated in the left and right panels, respectively.}
	\label{fig:Fitting:total_[N II]_and_[S II]}
\end{figure}

We explore the line ratios of [N II]/H$\alpha$, [S II]/H$\alpha$, and [S II]/[N
II] as functions of R$_{gal}$ in panels of the first and second rows in  
Figure~\ref{fig:Fitting:total_[N II]_and_[S II]}. The linear fitting results are summarized in Table \ref{R_fitting}.

For clarity in illustrating the trend, we divided the data into bins based on R$_{gal}$,
and display the median value along with 1$\sigma$ dispersion of each bin.
Interestingly, neither [N II]/H$\alpha$ nor [S II]/H$\alpha$ demonstrates a consistent, monotonous decrease with increasing R$_{gal}$. Instead, they peaks at
$\sim$ 9.1 kpc, within the interarm region between the Local arm and the Perseus arm. This peculiarity may be attributed to differences in the ionization
mechanisms between interarm regions and the arms, potentially due to a
substantial presence of HOLMES in the interarm region. Because of the non-monotonic trends, therefore, rather than attempting 
a least squares fitting of the data to derive possible slopes, we opt to simply 
connect the median values using a dashed line.

The line ratio of [S II]/[N II] is presented alongside R$_{gal}$ in the panel of the third row
in Figure~\ref{fig:Fitting:total_[N II]_and_[S II]}. It's found that [S II]/[N II]
increase with R$_{gal}$, exhibiting a a steep incline in the inner disk, and
maintaining nearly constancy in the outer disk. The disk is then divided into
inner and outer regions, using a cut-off at  R$_{gal}$ = 9.65 kpc. Linear fittings of
the trend in the inner disk yield with a slope of 0.1415 $\pm$ 0.0646 kpc$^{-1}$.

We examine the radial distributions of oxygen abundance in bottom two panels in Figure ~\ref{fig:Fitting:total_[N II]_and_[S II]}. 
The oxygen abundance demonstrates an inverse trend with R$_{gal}$ in the inner disk, with a slope of -0.0559 $\pm$ 0.0209 dex kpc$^{-1}$, while exhibiting a comparable pattern in the outer region, characterized by a slope of -0.0429 $\pm$ 0.0599 dex kpx$^{-1}$. A single linear fitting to the entire disk yields a slope of -0.0317 $\pm$ 0.0124 dex kpc$^{-1}$.

To explore the potential distribution differences between Sample-I and
Sample-II, we show the radial distribution of Sample-II in the right panels of
Figure~\ref{fig:Fitting:total_[N II]_and_[S II]}. It is found that [N
II]/H$\alpha$ gradually decreases with increasing R$_{gal}$. For the remaining
distributions, Sample-II exhibits similar trends to Sample-I. However, the slopes 
in the inner region are slightly steeper than that of Sample-I.

\subsection{Vertical distribution}

Based on WHAM data, \citet{Haffner_1999ApJ} have investigated the vertical 
distribution of DIG.  They found that the intensity of
H$\alpha$ emission decreases as distance from the Galactic midplane increases.
Additionally, the line ratios of [N II]/H$\alpha$ and [S II]/H$\alpha$ increase
with increasing $|z|$.

\begin{figure}[htbp]
	\centering

 	\includegraphics[scale=0.3]{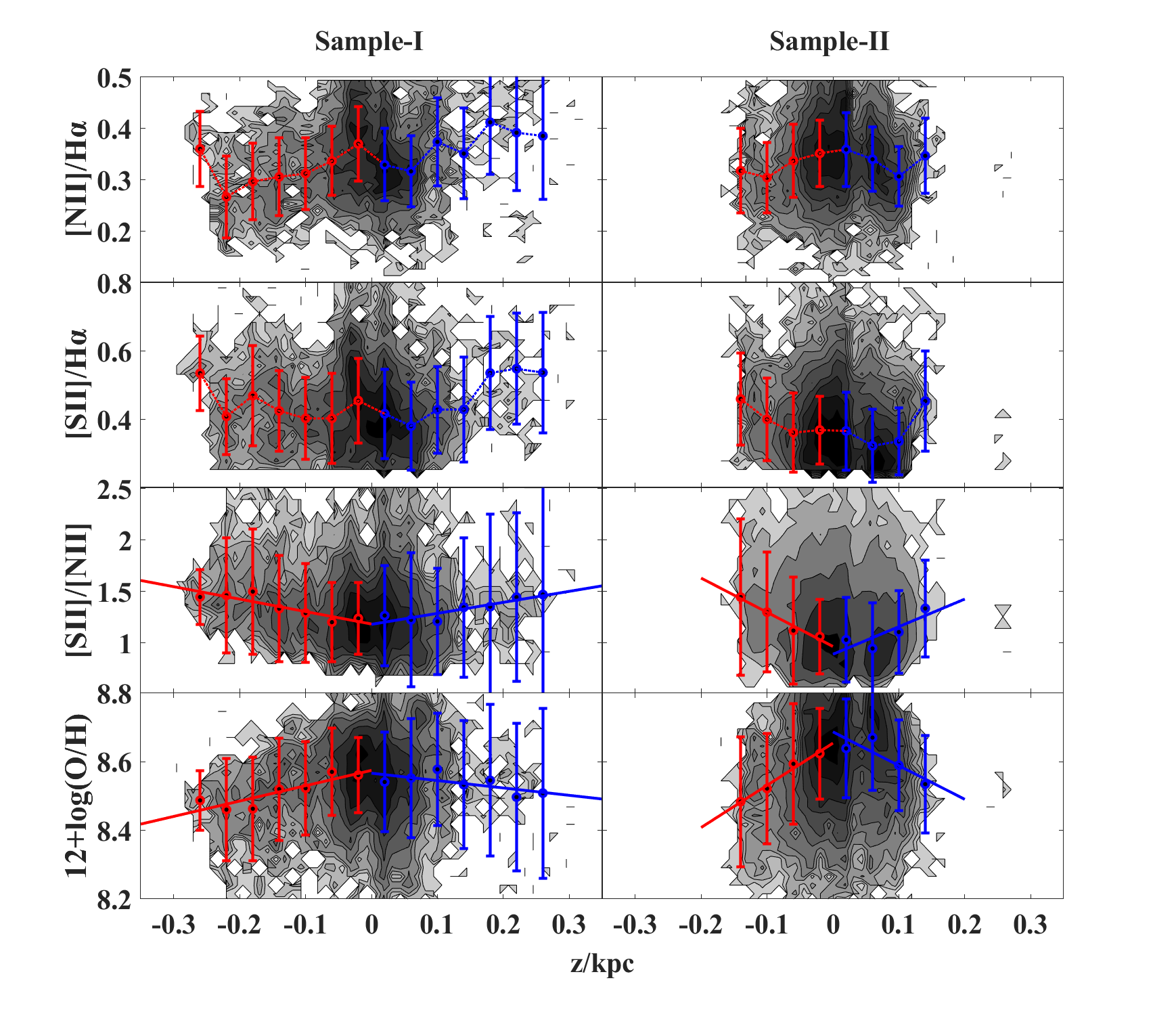}
	\caption{Similar to Figure~\ref{fig:Fitting:total_[N II]_and_[S II]}, with the variation plotted as a function of Galactic disk height (z).}
	\label{fig:Fitting:Z_distributions_removed_NGC2237}
\end{figure}

In Figure~\ref{fig:Fitting:Z_distributions_removed_NGC2237}, 
we examine the vertical distributions of line ratios and oxygen abundance of
Sample-I and Sample-II.  As these samples are located within a distance of less than 2.5 kpc from us, and the survey is constrained to $|b| \le 5^{\circ}$, they are predominantly situated in the thin disk. Sample-II, selected from regions adjacent to HII regions, is further restricted to even smaller heights from the midplane, with $|z| < $ 0.15 kpc, compared to $|z| < $ 0.3 kpc for Sample-I.

In the left panels in Figure ~\ref{fig:Fitting:Z_distributions_removed_NGC2237}, 
we examine the vertical distributions of line ratios and oxygen abundance of Sample-I. 
We confirmed that the [N II]/H$\alpha$ and [S II]/H$\alpha$ 
line ratios increase with increasing $|z|$. In comparison, the [S II]/[N II] ratio shows a more pronounced trend along the
vertical direction, with slopes of $1.2093 \pm 0.7067$ kpc$^{-1}$ and $1.0688 \pm 0.6047$ kpc$^{-1}$
in the southern and northern disks, respectively. Additionally, oxygen abundance is highest at the midplane and decreases with increasing $|z|$, with slopes of $-0.4476 \pm 0.2833$
dex kpc$^{-1}$ and $-0.2150 \pm 0.2640$ dex kpc$^{-1}$ in the southern and northern disks, respectively.

For comparison, Figure \ref{fig:Fitting:Z_distributions_removed_NGC2237} 
presents the results of Sample-II in its right panels. Compared to Sample-I, Sample-II 
shows comparable slopes for both the [S II]/[N II] ratio and oxygen abundance:
$3.3214 \pm 1.9115$ dex kpc$^{-1}$ and $2.6809 \pm 5.5085$ dex kpc$^{-1}$ for
[S II]/[N II] ratio in the southern and northern disks, respectively, and
$-1.2302 \pm 0.5969$ dex kpc$^{-1}$ and $-0.9834 \pm 1.8094$ dex kpc$^{-1}$
for oxygen abundance in these regions.

Detailed results of the linear fits are summarized in Table \ref{Z_fitting}. The differences in slopes between the southern and northern 
disks can be attributed to two main factors: 1) incomplete coverage of the
current sample across the disk, and 2) the separation of the southern and 
northern disk is dependent on R$_{gal}$ owing to the wrap structure of the
Galactic disk. In our upcoming work, we plan to further analyze these trends using a larger dataset that provides continuous coverage.

\section{Discussion}

\subsection{Oxygen Abundance}

Radial abundance gradients in the Galactic disk have been extensively investigated using various tracers, including open clusters, HII regions, PNe, and other stellar objects. In this work, the LAMOST MRS-N providess DIG as a new tracer to facilitate similar analyses and to compare the results with those obtained using HII regions.

In optical bands, oxygen abundances are accurately determined using the direct T$_e$-method, whereas the strong emission line method exhibits comparatively larger scatter. Conversely, in radio bands, T$_e$ can be measured with high precision ($\sim$ 1\%) using RRL and radio continuum, enabling the derivation of oxygen abundances through the tight oxygen-T$_e$ relationship.

Earlier studies by \citet{Balser_2011} determined an oxygen abundance gradient of -0.0446 $\pm$ 0.0049 dex kpc$^{-1}$ for 133 HII regions observed at National Radio Astronomy Observatory (NRAO). Among later works, \citet{Wenger_2019} observed radio emissions from 82 HII regions observed by  the Very Large Array (VLA) at NRAO, derived an oxygen abundance gradient of -0.052 $\pm$ 0.004 dex kpc$^{-1}$. \citet{Arellano-Cordova_2021} obtained an oxygen abundance gradient of 0.042 $\pm$ 0.009 dex kpc$^{-1}$ based on combined sample of 42 Galactic HII regions observed by the Gran Telescopio Canarias telescope and collected from public data. \citet{Mendez-Delgado_2022} recalibrated the distances to these 42 HII regions using Gaia EDR3 data, and derive the radial gradient for oxygen as -0.044 $\pm$ 0.009 dex kpc$^{-1}$. \citet{WangLiLi_2018} fitted the metallicity gradient of the HII regions of the whole Galactic disk and the outer disk (R$_{gal} \ge $ 11.5kpc) , calculated radial gradients of -0.036 $\pm$ 0.004 dex kpc$^{-1}$.

We adopt the N2S2H$\alpha$ method discussed in \citet{Dopita_2016}  to measure oxygen abundance for both Sample-I and Sample-II. In Sample-I, there is no evidence of a flattened gradient in the outer disk. Employing a single slope fitting yields an oxygen abundance radial gradient with a slope of -0.0317 $\pm$ 0.0124 dex kpc$^{-1}$, comparable to that observed in HII regions. In Sample-II, a clear two-slope pattern emerges, with the inner disk exhibiting a steeper slope than that of HII regions, while the outer disk shows nearly flat behavior.

It should be noted that the  N2S2H$\alpha$ method is calibrated based on assumptions about photoionization models. However, DIG encompasses various energy sources and cannot be accurately modeled solely by photoionization models. Therefore, applying the N2S2H$\alpha$ method to DIG may introduce unknown biases that need to be assessed using specific models designed for DIG in future studies.

\subsection{Variation of [S II]/[N II] ratio}

The ratio [N II]/H$\alpha$ in DIG regions is enhanced compared to that in HII regions, which can be attributed to the sensitivity of [N II]/H$\alpha$ to T$_e$ \citep{Haffner_2009}. According to the ionization potential of N  (14.5 eV) and N$^+$ (29.6 eV), it is implied
that most of N in the DIG region is likely to be in the form of N$^+$ and only a
small fraction of nitrogen exists as N$^{++}$ \citep{Madsen_2006ApJ}. On the other hand, the
ionization potential of S$^+$ is stated to be 23.4 eV, suggesting that a 
significant fraction of S may be present in the form of S$^{++}$. The quantity
[S II]/H$\alpha$ is described as a combination of T$_{e}$ and S$^+$/S.  The trend of [S II]/[N II]
versus H$\alpha$ intensity is reported to be relatively flat, implying that it is less
sensitive to changes in T$_e$ \citep{Haffner_1999ApJ}. Hereafter, the trend of [S II]/[N II] along the Galactic plane shown in Figure \ref{fig:Fitting:total_[N II]_and_[S II]} should not be caused by the increasing T$_{e}$ with R$_{gal}$.  Instead, it indicates that the ratio of S$^+$/S undergoes a systematic change with R$_{gal}$. That is, in the outer disk, a higher fraction of sulfur exists in the form of S$^{++}$.

\section{Summary}

We constructed a large DIG sample towards the anti-galactic center based on LAMOST MRS-N data and  investigate the radial and vertical distribution of  the line ratios, including [N II]/H$\alpha$, [S II]/H$\alpha$, [S II]/[N II], and oxygen abundance. Our main results are summarized as follows:

\begin{itemize} 

\item[(1)] The line ratios of [N II]/H$\alpha$ and [S II]/H$\alpha$ initially increase with R$_{gal}$ and then decrease, exhibiting a turnover point at approximately 9.1 kpc. This position corresponds to the interarm region, situated between the Local Arm and the Perseus Arm. This suggests a distinct ionizing mechanism between the DIG in the spiral arm and the interarm region.

\item[(2)] The [S II]/[N II] ratio initially rises with R$_{gal}$ in inner disk (8.34  kpc $<$ R$_{gal}$ $<$ 9.65 kpc) with a slope of 0.1415 $\pm$ 0.0646 kpc$^{-1}$, and then flatten out in the outer region, implying a higher ratio of S$^+$/S in the outer disk.

\item[(3)] The oxygen abundance is estimated using the N2S2H$\alpha$ method, which combines [S II]/[N II] and [N II]/H$\alpha$ ratios. The resulting oxygen abundance exhibits a comparable trend in both the inner and outer disk regions, showing slopes of -0.0559 $\pm$ 0.0209 dex kpc$^{-1}$ and -0.0429 $\pm$ 0.0599 dex kpc$^{-1}$, respectively. This trend can be described by a singular linear fit with a slope of -0.0317 $\pm$ 0.0124 dex kpc$^{-1}$.

\item[(4)] Despite the current sample being confined to the thin disk with $|z| < 0.3 $ kpc, we confirm that both [N II]/H$\alpha$ and [S II]/H$\alpha$  line ratios increase with increasing $|z|$ in both northern and southern disk. Additionally, [S II]/[N II] shows a more distinct trend, while the oxygen abundance exhibits an opposite trend.

\end{itemize} 

We possess a substantial sample of Galactic DIG. However, a notable limitation is the presence of substantial gaps between the observational plates, and these plates are not uniformly distributed across the Galactic disk. The asymmetrical distribution of DIG parameters found in our study between the southern and northern disks is likely due to constraints imposed by the current dataset and the the warp structure of the Galactic disk. To tackle these challenges, our plan is to improve our analysis by integrating additional survey data in our upcoming work.

\begin{acknowledgments}
 
This work is supported by the National Natural Science Foundation of China (NSFC) (No. 12090041; 12090040) and the National Key R\&D Program of China grant (No. 2021YFA1600401; 2021YFA1600400). This work is also sponsored by the Strategic Priority Research Program of the Chinese Academy of Sciences (No. XDB0550100). Guoshoujing Telescope (the Large Sky Area Multi-Object Fiber Spectroscopic Telescope LAMOST) is a National Major Scientific Project built by the Chinese Academy of Sciences. Funding for the project has been provided by the National Development and Reform Commission. LAMOST is operated and managed by the National Astronomical Observatories, Chinese Academy of Sciences.
                                                                                
\end{acknowledgments}

\bibliographystyle{aasjournal}

\bibliography{ref}

\begin{table}%[htbp]
	\centering
	\caption{Radial fitting results of each sample.}
	\begin{tabular}{ccccccc}
		\hline
		\multicolumn{2}{c}{\multirow{2}{*}{Samples}}&{\multirow{2}{*}{Parameters}}  & \multicolumn{4}{c}{Radial Gradient} \\

		\multicolumn{2}{c}{} && a(inner)(kpc$^{-1}$) & b(inner) & a(outer)(kpc$^{-1}$) & b(outer) \\
		\hline
		
		\multirow{3}{*}{Sample-I}&\multicolumn{2}{c}{[S II]/[N II]} 	&0.1415$\pm$0.0646&-0.0169$\pm$0.5806 & & \\

        \multirow{3}{*}{}&\multicolumn{2}{c}{12+log(O/H)(dex)} 	&-0.0559$\pm$0.0209&9.0528$\pm$0.1881&-0.0429$\pm$0.0599&8.9578$\pm$0.6093\\ 
        		\cline{1-7}
        \multirow{3}{*}{Sample-II}&\multicolumn{2}{c}{[S II]/[N II]} 	&0.2662$\pm$0.1147&-1.2174$\pm$1.0307& & \\

        \multirow{3}{*}{}&\multicolumn{2}{c}{12+log(O/H)(dex)} 	&-0.0994$\pm$0.0462&9.4631$\pm$0.4148&0.0559$\pm$0.0677&7.9815$\pm$0.6887\\ 

		\hline
	\end{tabular}
	
	\textbf{Notes}: The linear fitting function takes the form y = a $\times$ R$_{gal}$ + b. The inner disk is defined for 8.34 $<$ R$_{gal}$ $<$ 9.65 kpc, while the outer disk is indicated by R$_{gal}$ $>$ 9.65 kpc. 
	\label{R_fitting}
\end{table}

\begin{table}%[htbp]
	\centering
	\caption{Vertical fitting results of each sample.}%
	\begin{tabular}{ccccccc}
		\hline
		\multicolumn{2}{c}{\multirow{2}{*}{Samples}}&{\multirow{2}{*}{Parameters}}   &\multicolumn{4}{c}{Vertical Gradient}\\

		\multicolumn{2}{c}{} &&  a(south)(kpc$^{-1}$) & b(south) & a(north)(kpc$^{-1}$) & b(north)\\
		\hline
		
		\multirow{3}{*}{Sample-I}&\multicolumn{2}{c}{[S II]/[N II]} 	&1.2093$\pm$0.7067&1.1823$\pm$0.1139&1.0688$\pm$0.6047&1.1790$\pm$0.0975\\ 
		
        \multirow{3}{*}{}&\multicolumn{2}{c}{12+log(O/H)(dex)} 	&-0.4476$\pm$0.2833&8.5750$\pm$0.0457&-0.2150$\pm$0.2640&8.5667$\pm$0.0426\\ 
        		\cline{1-7}
        \multirow{3}{*}{Sample-II}&\multicolumn{2}{c}{[S II]/[N II]} 	 &3.3214$\pm$1.9115&0.9644$\pm$0.1752&2.6809$\pm$5.5085&0.8884$\pm$0.5049\\ 
        
        \multirow{3}{*}{}&\multicolumn{2}{c}{12+log(O/H)(dex)} 	&-1.2302$\pm$0.5969&8.6543$\pm$0.0547&-0.9834$\pm$1.8094&8.6872$\pm$0.1658\\ 

		\hline
	\end{tabular}
	
	\textbf{Notes}:  The linear fitting function takes the form y = a $\times$ $|$z$|$ + b.  The northern and southern disks correspond to z $>$ 0 and z $<$ 0, respectively. 
	\label{Z_fitting}
\end{table}

\end{document}